\documentclass[10pt]{article}
\usepackage{amsfonts,amsmath,amssymb,graphicx}
\begin{document}
\title{Alternative construction of the closed form of the Green's
function for the wavized Maxwell fish-eye problem}
\author{Rados{\l}aw Szmytkowski \\*[3ex]
Atomic Physics Division,
Department of Atomic Physics and Luminescence, \\
Faculty of Applied Physics and Mathematics,
Gda{\'n}sk University of Technology, \\
Narutowicza 11/12, 80--233 Gda{\'n}sk, Poland \\
email: radek@mif.pg.gda.pl}
\date{\today}
\maketitle
\begin{abstract}
In the recent paper [J.\ Phys.\ A 44 (2011) 065203], we have arrived
at the closed-form expression for the Green's function for the
partial differential operator describing propagation of a scalar wave
in an $N$-dimensional ($N\geqslant2$) Maxwell fish-eye medium. The
derivation has been based on unique transformation properties of the
fish-eye wave equation under the hyperspherical inversion. In this
communication, we arrive at the same expression for the fish-eye
Green's function following a different route. The alternative
derivation we present here exploits the fact that there is a close
mathematical relationship, through the stereographic projection,
between the wavized fish-eye problem in $\mathbb{R}^{N}$ and the
problem of propagation of scalar waves over the surface of the
$N$-dimensional hypersphere.
\end{abstract}
%
%
In the recent paper \cite{Szmy11}, we have constructed the
closed-form expression for the Green's function for the partial
differential operator describing propagation of a scalar wave in an
$N$-dimensional ($N\geqslant2$) Maxwell fish-eye medium. Our
considerations, inspired by an earlier work of Demkov and Ostrovsky
\cite{Demk71}, have been based on the use of unique transformation
properties of the scalar fish-eye wave equation under the
hyperspherical inversion. In this communication, we show it is
possible to arrive at the same representation of the fish-eye Green's
function proceeding along a different but, we believe, equally
elegant route. The reasoning we present below is conceptually rooted
in the brilliant observation made several decades ago by
Carath{\'e}odory \cite{Cara26}, who pointed out, in the context of
geometrical optics, that the remarkable properties of the Maxwell
fish-eye are related to the one-to-one stereographic-projection
correspondence between propagation in that medium and the free motion
on the sphere (cf also Refs.\ \cite{Fran90,Fran91}).

\begin{figure}
\begin{center}
\includegraphics{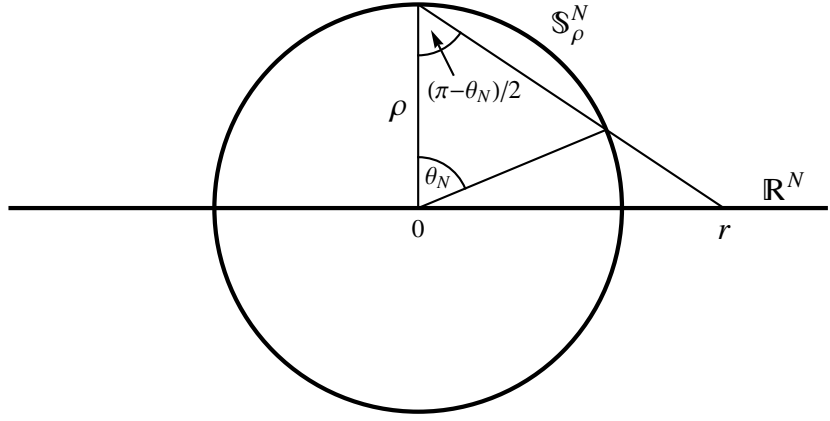}
\end{center}
\caption{The transformation (\ref{3}) is the inverse stereographic 
projection of the space $\mathbb{R}^{N}$ onto the hypersphere 
$\mathbb{S}_{\rho}^{N}$ of radius $\rho$.}
\label{fig1}
\end{figure}

To begin, we observe that the fish-eye Green's function in
$\mathbb{R}^{N}$, $N\geqslant2$, solves the inhomogeneous partial
differential equation
\begin{equation}
\left[\boldsymbol{\nabla}_{\mathbb{R}^{N}}^{2}
+\frac{4\nu(\nu+1)\rho^{2}}{(r^{2}+\rho^{2})^{2}}\right]
G_{\nu}(\boldsymbol{r},\boldsymbol{r}')
=\delta^{(N)}(\boldsymbol{r}-\boldsymbol{r}'),
\label{1}
\end{equation}
where $\boldsymbol{\nabla}_{\mathbb{R}^{N}}^{2}$ is the Laplace
operator in $\mathbb{R}^{N}$ with respect to coordinates of the
observation point $\boldsymbol{r}$, $\boldsymbol{r}'$ is the point
where the unit delta source is located, $\rho>0$ and
$\nu\in\mathbb{C}$. After introducing the hyperspherical coordinates
$\{r,\Omega_{N-1}\}$, with $r=|\boldsymbol{r}|$ and with
$\Omega_{N-1}$ standing collectively for $N-1$ angles characterizing
the orientation of the radius vector $\boldsymbol{r}$ (and similarly
for $\boldsymbol{r}'$), Eq.\ (\ref{1}) casts into the form
\begin{equation}
\left[\frac{\partial^{2}}{\partial r^{2}}
+\frac{N-1}{r}\frac{\partial}{\partial r}
+\frac{1}{r^{2}}\boldsymbol{\nabla}_{\mathbb{S}^{N-1}}^{2}
+\frac{4\nu(\nu+1)\rho^{2}}{(r^{2}+\rho^{2})^{2}}\right]
G_{\nu}(\boldsymbol{r},\boldsymbol{r}')=\frac{\delta(r-r')
\delta^{(N-1)}(\Omega_{N-1}-\Omega_{N-1}^{\prime})}
{r^{(N-1)/2}r^{\prime\,(N-1)/2}},
\label{2}
\end{equation}
where $\boldsymbol{\nabla}_{\mathbb{S}^{N-1}}^{2}$ is the
Laplace--Beltrami operator on the unit hypersphere
$\mathbb{S}^{N-1}$. Now we make the most crucial step in our
reasoning and switch from the radial variables $r$ and $r'$ to the
angular variables $\theta_{N}$ and $\theta_{N}^{\prime}$, according
to
\begin{equation}
\cot\frac{\theta_{N}}{2}=\frac{r}{\rho},
\qquad
\cot\frac{\theta_{N}^{\prime}}{2}=\frac{r'}{\rho}
\qquad (0\leqslant\theta_{N},\theta_{N}^{\prime}\leqslant\pi),
\label{3}
\end{equation}
the angular coordinate ensembles $\Omega_{N-1}$ and
$\Omega_{N-1}^{\prime}$ remaining unchanged. The geometrical meaning
of the transformation (\ref{3}) becomes obvious after a glance at
Fig.\ \ref{fig1}: this is the inverse stereographic projection of the
space $\mathbb{R}^{N}$ onto the hypersphere $\mathbb{S}_{\rho}^{N}$
of radius $\rho$, the space to be projected being the equatorial
hyperplane of the hypersphere.
Since, in view of Eq.\ (\ref{3}) and of the well-know properties of
the Dirac delta, it holds that
\begin{equation}
\delta(r-r')=\frac{2}{\rho}\sin\frac{\theta_{N}}{2}
\sin\frac{\theta_{N}^{\prime}}{2}\,
\delta(\theta_{N}-\theta_{N}^{\prime}),
\label{4}
\end{equation}
the transformation in question changes Eq.\ (\ref{2}) into
\begin{eqnarray}
&& \left[\frac{\partial^{2}}{\partial\theta_{N}^{2}}
+\left(\cot\frac{\theta_{N}}{2}-\frac{N-1}{\sin\theta_{N}}\right)
\frac{\partial}{\partial\theta_{N}}
+\frac{\boldsymbol{\nabla}_{\mathbb{S}^{N-1}}^{2}}
{\sin^{2}\theta_{N}}+\nu(\nu+1)\right]
G_{\nu}(\boldsymbol{r},\boldsymbol{r}')
\nonumber \\
&& \hspace*{10em} =\,\frac{1}{2\rho^{N-2}}
\frac{\delta(\theta_{N}-\theta_{N}^{\prime})
\delta^{(N-1)}(\Omega_{N-1}-\Omega_{N-1}^{\prime})}
{\sin\frac{\theta_{N}}{2}\cot^{(N-1)/2}\frac{\theta_{N}}{2}
\sin\frac{\theta_{N}^{\prime}}{2}
\cot^{(N-1)/2}\frac{\theta_{N}^{\prime}}{2}}.
\label{5}
\end{eqnarray}
Both the differential operator on the left-hand side and the
multiplier of the deltas on the right-hand side of Eq.\ (\ref{5})
look complicated. However, a remarkable simplification is achieved
after one replaces the Green's function
$G_{\nu}(\boldsymbol{r},\boldsymbol{r}')$ by the function
$\mathcal{G}_{\nu-N/2+1}(\Omega_{N},\Omega_{N}^{\prime})$, the two
being related by
\begin{equation}
G_{\nu}(\boldsymbol{r},\boldsymbol{r}')
=\left(\frac{2}{\rho}\right)^{N-2}\sin^{N-2}\frac{\theta_{N}}{2}
\sin^{N-2}\frac{\theta_{N}^{\prime}}{2}\,
\mathcal{G}_{\nu-N/2+1}(\Omega_{N},\Omega_{N}^{\prime}).
\label{6}
\end{equation}
Here, $\Omega_{N}$ stands for the set $\{\theta_{N},\Omega_{N-1}\}$
(and similarly for $\Omega_{N}^{\prime}$); the reason for attaching
the particular subscript to $\mathcal{G}$ will become clear shortly.
Insertion of Eq.\ (\ref{6}) into Eq.\ (\ref{5}), followed by some
obvious rearrangements, results in
\begin{eqnarray}
&& \left[\frac{\partial^{2}}{\partial\theta_{N}^{2}}
+(N-1)\cot\theta_{N}\frac{\partial}{\partial\theta_{N}}
+\frac{\boldsymbol{\nabla}_{\mathbb{S}^{N-1}}^{2}}{\sin^{2}\theta_{N}}
+\left(\nu-\frac{N}{2}+1\right)\left(\nu+\frac{N}{2}\right)\right]
\mathcal{G}_{\nu-N/2+1}(\Omega_{N},\Omega_{N}^{\prime})
\nonumber \\
&& \hspace*{20em} =\,\frac{\delta(\theta_{N}-\theta_{N}^{\prime})
\delta^{(N-1)}(\Omega_{N-1}-\Omega_{N-1}^{\prime})}
{\sin^{(N-1)/2}\theta_{N}\sin^{(N-1)/2}\theta_{N}^{\prime}}.
\label{7}
\end{eqnarray}
The first three terms in the square bracket on the left-hand side of
Eq.\ (\ref{7}) are immediately recognized to form the
Laplace--Beltrami operator on the unit hypersphere $\mathbb{S}^{N}$:
\begin{equation}
\frac{\partial^{2}}{\partial\theta_{N}^{2}}
+(N-1)\cot\theta_{N}\frac{\partial}{\partial\theta_{N}}
+\frac{\boldsymbol{\nabla}_{\mathbb{S}^{N-1}}^{2}}{\sin^{2}\theta_{N}}
\equiv\boldsymbol{\nabla}_{\mathbb{S}^{N}}^{2}
\qquad (N\geqslant2),
\label{8}
\end{equation}
while the expression on the right-hand side of Eq.\ (\ref{7}) is
simply the Dirac delta on $\mathbb{S}^{N}$:
\begin{equation}
\frac{\delta(\theta_{N}-\theta_{N}^{\prime})
\delta^{(N-1)}(\Omega_{N-1}-\Omega_{N-1}^{\prime})}
{\sin^{(N-1)/2}\theta_{N}\sin^{(N-1)/2}\theta_{N}^{\prime}}
=\delta^{(N)}(\Omega_{N}-\Omega_{N}^{\prime}).
\label{9}
\end{equation}
Hence, with the definition
\begin{equation}
\lambda=\nu-\frac{N}{2}+1,
\label{10}
\end{equation}
Eq.\ (\ref{7}) may be rewritten compactly as
\begin{equation}
\left[\boldsymbol{\nabla}_{\mathbb{S}^{N}}^{2}
+\lambda(\lambda+N-1)\right]
\mathcal{G}_{\lambda}(\Omega_{N},\Omega_{N}^{\prime})
=\delta^{(N)}(\Omega_{N}-\Omega_{N}^{\prime}).
\label{11}
\end{equation}
This is the equation defining the Green's function for the Helmholtz
operator on the hypersphere $\mathbb{S}^{N}$; it has been studied by
us in Ref.\ \cite{Szmy07}. There, it has been shown that the solution
to Eq.\ (\ref{11}) is
\begin{equation}
\mathcal{G}_{\lambda}(\Omega_{N},\Omega_{N}^{\prime})
=\frac{\pi C_{\lambda}^{(N-1)/2}
\left(-\cos\angle(\Omega_{N},\Omega_{N}^{\prime})\right)}
{(N-1)S_{N}\sin(\pi\lambda)},
\label{12}
\end{equation}
where $C_{\lambda}^{\alpha}(\xi)$ is the Gegenbauer function,
$\angle(\Omega_{N},\Omega_{N}^{\prime})$ is the angle between the
directions $\Omega_{N}$ and $\Omega_{N}^{\prime}$, while
\begin{equation}
S_{N}=\frac{2\pi^{(N+1)/2}}{\Gamma\left(\frac{N+1}{2}\right)}
\label{13}
\end{equation}
is the area of $\mathbb{S}^{N}$. Hence, on invoking Eq.\ (\ref{6}),
we see that the closed-form representation of the fish-eye Green's
function in $\mathbb{R}^{N}$ is
\begin{equation}
G_{\nu}(\boldsymbol{r},\boldsymbol{r}')
=\frac{2^{N-4}\Gamma\left(\frac{N-1}{2}\right)}
{\rho^{N-2}\pi^{(N-1)/2}
\sin\left[\pi\left(\frac{N}{2}-\nu\right)\right]}
\sin^{N-2}\frac{\theta_{N}}{2}
\sin^{N-2}\frac{\theta_{N}^{\prime}}{2}\,C_{\nu-N/2+1}^{(N-1)/2}
\left(-\cos\angle(\Omega_{N},\Omega_{N}^{\prime})\right).
\label{14}
\end{equation}
To accomplish the task fully, we have to express the right-hand side
of Eq.\ (\ref{14}) in terms of the radius vectors $\boldsymbol{r}$
and $\boldsymbol{r}'$ instead of the hyperangles $\Omega_{N}$ and
$\Omega_{N}^{\prime}$. To this end, at first we observe that the
cosine of the angle $\angle(\Omega_{N},\Omega_{N}^{\prime})$ may be
written as
\begin{equation}
\cos\angle(\Omega_{N},\Omega_{N}^{\prime})
=\cos\theta_{N}\cos\theta_{N}^{\prime}
+\sin\theta_{N}\sin\theta_{N}^{\prime}
\cos\angle(\Omega_{N-1},\Omega_{N-1}^{\prime}).
\label{15}
\end{equation}
However, from Eq.\ (\ref{3}) it follows that
\begin{equation}
\cos\theta_{N}=\frac{\cot^{2}\frac{\theta_{N}}{2}-1}
{\cot^{2}\frac{\theta_{N}}{2}+1}
=\frac{r^{2}-\rho^{2}}{r^{2}+\rho^{2}}
\label{16}
\end{equation}
and
\begin{equation}
\sin\theta_{N}=\frac{2\cot\frac{\theta_{N}}{2}}
{\cot^{2}\frac{\theta_{N}}{2}+1}=\frac{2\rho r}{r^{2}+\rho^{2}}
\label{17}
\end{equation}
(and similarly for $\cos\theta_{N}^{\prime}$ and
$\sin\theta_{N}^{\prime}$), so that
\begin{eqnarray}
\cos\angle(\Omega_{N},\Omega_{N}^{\prime})
&=& 1-\frac{2\rho^{2}[r^{2}+r^{\prime\,2}
-2rr'\cos\angle(\Omega_{N-1},\Omega_{N-1}^{\prime})]}
{(r^{2}+\rho^{2})(r^{\prime\,2}+\rho^{2})}
\nonumber \\
&=& 1-\frac{2\rho^{2}(\boldsymbol{r}-\boldsymbol{r}')^{2}}
{(r^{2}+\rho^{2})(r^{\prime\,2}+\rho^{2})}.
\label{18}
\end{eqnarray}
Furthermore, invoking Eq.\ (\ref{3}) again, we see that
\begin{equation}
\sin\frac{\theta_{N}}{2}=\frac{1}
{\sqrt{\cot^{2}\frac{\theta_{N}}{2}+1}}
=\frac{\rho}{\sqrt{r^{2}+\rho^{2}}}
\label{19}
\end{equation}
(and similarly for $\sin\frac{\theta_{N}^{\prime}}{2}$). Plugging
Eqs.\ (\ref{18}) and (\ref{19}) into Eq.\ (\ref{14}), we eventually
arrive at 
\begin{equation}
G_{\nu}(\boldsymbol{r},\boldsymbol{r}')
=\frac{2^{N-4}\Gamma\left(\frac{N-1}{2}\right)}
{\pi^{(N-1)/2}\sin\left[\pi\left(\frac{N}{2}-\nu\right)\right]}
\frac{\displaystyle\rho^{N-2}C_{\nu-N/2+1}^{(N-1)/2}
\left(-1+\frac{2\rho^{2}(\boldsymbol{r}-\boldsymbol{r}')^{2}}
{(r^{2}+\rho^{2})(r^{\prime\,2}+\rho^{2})}\right)}
{(r^{2}+\rho^{2})^{N/2-1}(r^{\prime\,2}+\rho^{2})^{N/2-1}}.
\label{20}
\end{equation}
This representation of the fish-eye Green's function in
$\mathbb{R}^{N}$ is identical with the one found by us in Ref.\
\cite[Eq.\ (3.42)]{Szmy11} using the hyperspherical inversion
technique.
%
%

%
\end{document}